\documentstyle[aps,twocolumn]{revtex}
\tighten
\begin{document}
\draft
\newcommand{\ie}{{\em i.e.}}
\newcommand{\eg}{{\em e.g.}}
\newcommand{\vs}{{\em vs.}}
\title{
KINETIC ISING SYSTEMS AS MODELS OF \\
MAGNETIZATION SWITCHING IN
SUBMICRON FERROMAGNETS
}

\author{
Howard~L.~Richards$^{* \dag}$,
Scott~W.~Sides$^{* \dag}$,
Mark~A.~Novotny$^{\dag}$, and
Per~Arne~Rikvold$^{* \dag}$
}
\address{
$^*$Center for Materials Research and Technology
and Department of Physics \\
$^{\dag}$Supercomputer Computations Research Institute \\
Florida State University,
Tallahassee, FL 32306-3016, USA
}
\date{\today}
\maketitle

\begin{abstract}
Recently experimental techniques, such as magnetic force microscopy
(MFM), have enabled the magnetic state of individual sub-micron
particles to be resolved.
Motivated by these experimental developments,
we use Monte Carlo simulations of two-dimensional kinetic
Ising ferromagnets to study the magnetic relaxation
in a negative applied field of a grain with an initial
magnetization $m_0 \! = \! +1$.
The magnetostatic dipole-dipole interactions
are treated to lowest order by adding to the Hamiltonian a
term proportional to the square of the magnetization.
We use droplet theory to predict the functional
forms for some quantities which can be observed by MFM.
One such quantity is the probability
that the magnetization is positive,
which is a function of time, field, grain size, and grain
dimensionality.
The relaxation is characterized by the number of droplets larger
than a field-dependent critical size which form during the switching
process.
Our simulations of the kinetic Ising model
are in excellent agreement with droplet-theoretical predictions.
The qualitative agreement between experiments and our simulations
of switching in individual single-domain ferromagnets
indicates that the switching mechanism in such particles may involve
local nucleation and subsequent growth of droplets of the stable
phase.
\end  {abstract}

\section{Introduction}

The processes by which magnetization reversal occurs in the
nanoscale ferromagnets that will make up the next-generation
recording media
are the subject of active research.
One quantity for which theory
and experiment often disagree is the lifetime
$\tau$, which is the time
required for a particle with initial
magnetization $m_0 \! = \! +1$
to reach $m \! = \! 0$ when a
magnetic field in the $-\hat{z}$ direction is applied.
Micromagnetics,\cite{WFBrown} a theoretic technique
in which differential equations are numerically solved
on a coarse-grained lattice, predicts that the lifetime is given
by the Arrhenius equation
\begin{equation}
  \label{eq:Arrhenius}
	\tau \propto \exp (\beta \Delta F)
\end  {equation}
with $\Delta F \! \propto \! L^d$.
Here $\beta^{-1}$ is the temperature in units of energy,
$\Delta F$ is the free-energy barrier between the stable and
metastable phases, and $L$ is the linear system size.
This same prediction is made by the
standard N{\' e}el-Brown theory of single-domain
ferromagnets.\cite{Neel49,Brown}
The evident failure of Eq.~(\ref{eq:Arrhenius}) with
$\Delta F \! \propto \! L^d$ for somewhat larger grains is
ascribed to the existence of more than one domain in larger
particles, as is a corresponding peak in plots
\vs\ $L$ of
the switching field $H_{\rm sw}$, which is
the field required to yield a given lifetime.

Recently techniques such as MFM have
been used to resolve the magnetic properties of
{\em isolated, well-characterized} single-domain particles
(see, \eg, Ref.~\cite{MFM}).
This is an important advance, since
previous experiments on ferromagnetic powders left
uncertainties due to the range of grain sizes and orientations
and the local magnetic environments.  Observations of individual
particles by MFM have made it clear that even
for many {\em single-domain}
particles, N{\' e}el-Brown theory is inadequate.

We have applied the statistical-mechanical droplet theory of
metastable decay to nanoscale ferromagnets with large uniaxial
anisotropy, and compared Monte Carlo simulations of
square-lattice Ising systems with droplet-theory
predictions.\cite{2dpi}
(For a review of droplet theory, see Ref.~\cite{RikARCP94}.)
The agreement between theory and simulation is quite good,
and despite the crudeness of the Ising model as a model for
real magnets, it shows good qualitative agreement with the
MFM experiments.
We find rich $L$-dependent behavior in the standard Ising model,
even though its equilibrium structure is a single domain
for all $L$.
This suggests that for some strongly anisotropic magnetic
materials, magnetization reversal may occur through the
nucleation and growth of non-equilibrium droplets.
Details of our work are given in Refs.~\cite{2dpi} and
\cite{demag}, including
formulae for general dimensionality.
For simplicity, we only discuss the two-dimensional case
here.

\section{Applied Droplet Theory}

To be concrete, consider a two-dimensional kinetic Ising
ferromagnet ($s_i \! = \! \pm 1$) with Hamiltonian
\begin{equation}
  \label{eq:hamiltonian}
	{\cal H} = -J \sum_{\rm n.n.} s_i s_j - H \sum_i s_i
	           + L^{-2} D \biggl( \sum_i s_i \biggr)^2
\end  {equation}
and Metropolis single-spin-flip dynamics on a square lattice
with periodic boundary conditions.
The last term represents a
mean-field approximation for the dipole-dipole interaction
energy and is taken to be zero except as noted.
The critical radius of a ``droplet'' of $s_i \! = \! -1$ spins
surrounded by $s_i \! = \! +1$ spins occurs when the
free energy of the droplet
($2 \pi R \sigma \! - \! \pi R^2 2|H|$)
is maximum:
\(
	R_c \!\approx \! \sigma / 2|H|,
\)
where $\sigma$ is the surface tension per unit length.
Droplets smaller than this will very probably shrink and
vanish; larger droplets will very probably grow
and reverse the magnetization of the system.
In a sufficiently large system, the probability per unit time
that a critical droplet forms, centered at a given site, is
given by droplet theory as \cite{RikARCP94}
\begin{equation}
  \label{eq:NucRate}
   I \propto |H|^3 \exp \left( - \beta \pi \sigma^2 / 2 |H|
			\right) \ .
\end  {equation}
The details of the magnetization reversal depends on the number of
critical droplets the system forms.

\begin{figure}
\vspace*{2.7truein}
	 \includegraphics{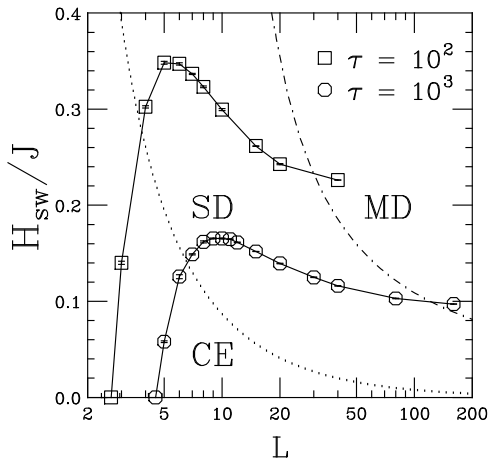}
\caption{ 	 The relation between the switching field
		 $H_{\rm sw}$ and the system
		 width $L$ for two different fixed lifetimes
		 (solid curves), calculated
		 by kinetic Ising model simulations
		 at $k_{\rm B}T \! = \! 0.8 k_{\rm B}T_{\rm c}
			  \! \approx \! 1.815 J.$
		 The dotted curve is near the crossover between the
		 CE and SD regions;\protect\cite{2dpi}
		 the dash-dotted curve is
		 near the crossover between the SD and MD regions.
		 \protect\cite{2dpi} }
\label{fig:H_sw}
\end{figure}

For weak fields or small systems ($L \! < \! R_c$),
no critical droplet
can form.  This is called the Coexistence (CE) Region, and since
two interfaces (remember: periodic boundary conditions)
must form to reverse the magnetization,
\begin{equation}
  \label{eq:CELife}
   \tau \propto \exp \left\{ \beta \left[ 2 \sigma L -
			O\left(HL^2\right) \right] \right\} \ .
\end  {equation}
For slightly larger $L$, the first supercritical droplet will
grow to the size of the system before another one can form.
The lifetime in this Single Droplet (SD) Region is
\begin{equation}
  \label{eq:SDLife}
   \tau \approx \left[L^d I \right]^{-1} \ .
\end  {equation}
In both the CE and SD regions, switching is a
Poisson process, so the standard deviation of the
lifetime is comparable to $\tau$.
Both Eq.~(\ref{eq:CELife}) and Eq.~(\ref{eq:SDLife}) are actually
special cases of the Arrhenius equation, Eq.~(\ref{eq:Arrhenius}),
but in neither case is $\Delta F$ proportional to $L^d$.
Note that if $\tau$ is held constant and the system size is
increased, Eq.~(\ref{eq:CELife}) implies that the magnetic field
must {\em increase}, whereas Eq.~(\ref{eq:SDLife}) implies that
the magnetic field must {\em decrease}.  This shows that
the peak in $H_{\rm sw}$
occurs near the crossover between the
CE and SD regions (see Fig.~\ref{fig:H_sw}).

\begin{figure}
\vspace*{2.7truein}
	 \includegraphics{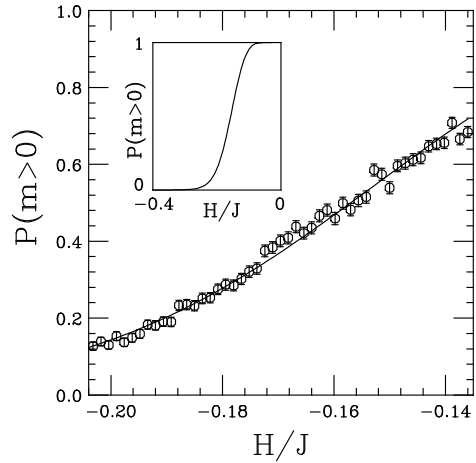}
\caption{ 	 The probability that $m \! > \! 0$
		 for a kinetic Ising
		 system in the SD region.
		 $T \! = \! 0.8 T_c$,
		 $t \! = \! 914$ Monte Carlo steps per spin
                     (MCSS) and
		     $L \! = \! 10$.
		     The solid curve is a fit of
		     $\exp(-t/\tau)$ to the MC data, where
		     $\tau$ is given by
		     Eq.~(\protect\ref{eq:SDLife}).
		     The inset figure shows the fitted curve over
		     a wider range in $H$.
}
\label{fig:SD_P}
\end{figure}

\begin{figure}
\vspace*{2.7truein}
	 \includegraphics{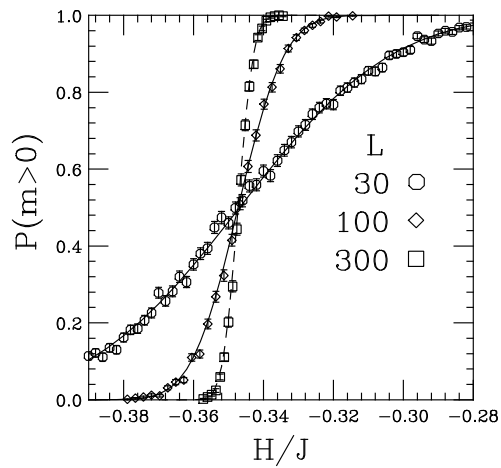}
\caption{ 	 The probability that $m \! > \! 0$
		 for a kinetic Ising
		 system in the MD region.
		 $T \! = \! 0.8 T_c$,
		 $\tau \! = \! 40.7$ MCSS and
		     $L \! = \! 30$, 100, and 300.
		     The solid curves are fits of
		     droplet theory predictions
		     to the MC data.
		     The dashed curve is the fit of
		     the droplet-theory prediction
		     for $L \! = \! 100$
		     extrapolated to $L \! = \! 300$.
}
\label{fig:MD_P}
\end{figure}

The probability that the magnetization is greater than zero
$P(m \! > \!0)$
is shown as a function of
field in Fig.~\ref{fig:SD_P} for a system in the SD region.
This probability is what is most easily observed in MFM experiments,
and it decays exponentially with time in both the CE and SD regions.
In the SD region, the system is very unlikely to return to the the
metastable state from the stable state, so
\begin{equation}
  \label{eq:SD_P_t}
	P(m \! > \!0) = \exp \left(-t/\tau \right) \; .
\end  {equation}
In the CE region such backwards switching
takes place on a timescale comparable with
the initial decay, so the situation is more complicated.

For sufficiently large $L$ or $H$,
several supercritical droplets may form
before any one of them has grown to the size of the system.
This is the Multi-Droplet Region (MD).
Such systems were first studied by Kolmogorov,\cite{Kolmogorov37}
Johnson and Mehl,\cite{JohnsMehl39}
and Avrami,\cite{Avrami} and have a lifetime
\begin{equation}
  \label{eq:MDLife}
	\tau \approx \left[ I\pi v^2/3 \ln 2 \right]^{-1/3} \ ,
\end  {equation}
where the radial growth velocity $v$ is assumed to be proportional
to $|H|$.  Although $\tau$ is independent of $L$,
the variance of the lifetime is proportional to $(v/L)^2$.
Measuring $P(m \! > \! 0)$ as a function of
$H$ or $t$ thus provides a means of estimating the proportionality
constant between $v$ and $H$
(see Fig.~\ref{fig:MD_P}).  Details are given in Ref.~\cite{2dpi}.

\begin{figure}
\vspace*{2.7truein}
	 \includegraphics{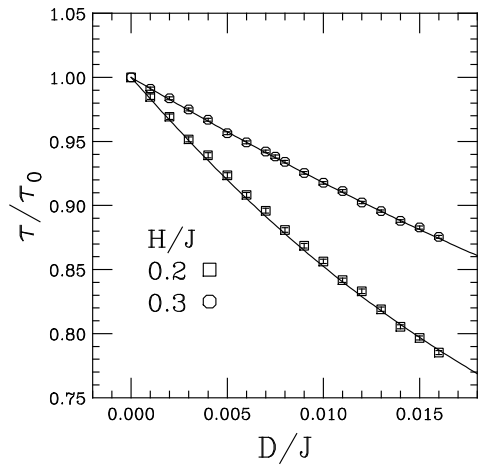}
\caption{ 	 The lifetime $\tau$ for a kinetic Ising
		 system in the MD region, normalized by the
		 lifetime in a similar system with $D \! = \! 0$.
		 $T \! = \! 0.8 T_c$.
		     The solid curves are
		     droplet-theory predictions.
}
\label{fig:md_tau_D}
\end{figure}

The addition of the dipole-dipole interaction energy in
Eq.~(\ref{eq:hamiltonian}) makes the forms for the lifetime
somewhat more complicated than we have presented here.
In the CE and SD regions, a form of the Arrhenius equation
[Eq.~(\ref{eq:Arrhenius})] still applies, and
switching is still a Poisson process, so
Eq.~(\ref{eq:SD_P_t}) still applies for the SD region.
In the MD region, the system evolves in a time-dependent
effective field, $H_{\rm eff} \! \equiv \! H - 2Dm(t)$.
The time-dependent magnetization can be found to $O(D)$
fairly easily, and we find numerically that the
$O(D^2)$ correction is relatively small.  We can then
solve analytically for $\tau$ to $O(D^2)$ and find
good agreement with simulation results
(Fig.~\ref{fig:md_tau_D}).  A detailed treatment of
the $D \! > \! 0$ case is given in Ref.~\cite{demag}.

\section{Conclusion}

We have used Monte Carlo methods to simulate magnetization
switching in two-dimensional kinetic Ising ferromagnets.
The results of the simulations can be well explained by
droplet theory and show good qualitative agreement with
experiments, despite the comparative simplicity of the Ising model.
This simplicity, in turn, allows us to develop an understanding
of the underlying statistical mechanics.  Particular features
to make the model more realistic, such as appropriate boundary
conditions, quenched randomness, and less rigorous anisotropy
will be added in later studies.

\acknowledgments

We thank S.~von Moln{\' a}r, D.~M.\ Lind,
J.~W.\ Harrell, and W.~D.\ Doyle for useful discussions.
This research was supported in part by
FSU-MARTECH,
by
FSU-SCRI
under DOE
Contract No.\ DE-FC05-85ER25000,
and by
NSF
Grants No.\ DMR-9315969 and DMR-9520325.



\end{document}